\documentclass[12pt]{article}
\usepackage{pic02}
\usepackage{hyperref}
\usepackage{url}
\usepackage{graphicx}

\begin{document}

\title{\bf NEUTRINO PHYSICS AND ASTRONOMY WITH MACRO}
\author{P. Bernardini, for the MACRO Collaboration                    \\
{\em Dipartimento di Fisica dell'Universit\`a and INFN, Lecce, Italy} \\
}
\maketitle


\baselineskip=14.5pt
\begin{abstract}
MACRO experiment operated in the Gran Sasso underground laboratory. 
Neutrino events collected by this detector are used in order to study 
the atmospheric neutrino flux. Different measurements in different 
energy samples are in full agreement and show evidence of neutrino 
oscillation phenomenon with maximal mixing and $\Delta m^2 \sim 0.0025\ eV^2$. 
Also the search for neutrino astrophysical sources is reported.
\end{abstract}

\baselineskip=17pt

\section{MACRO as an atmospheric neutrino detector}

\begin{figure}[t]
  \centerline{\hbox{ 
    \hspace{-0.5cm}
    \includegraphics[width=6.6cm]{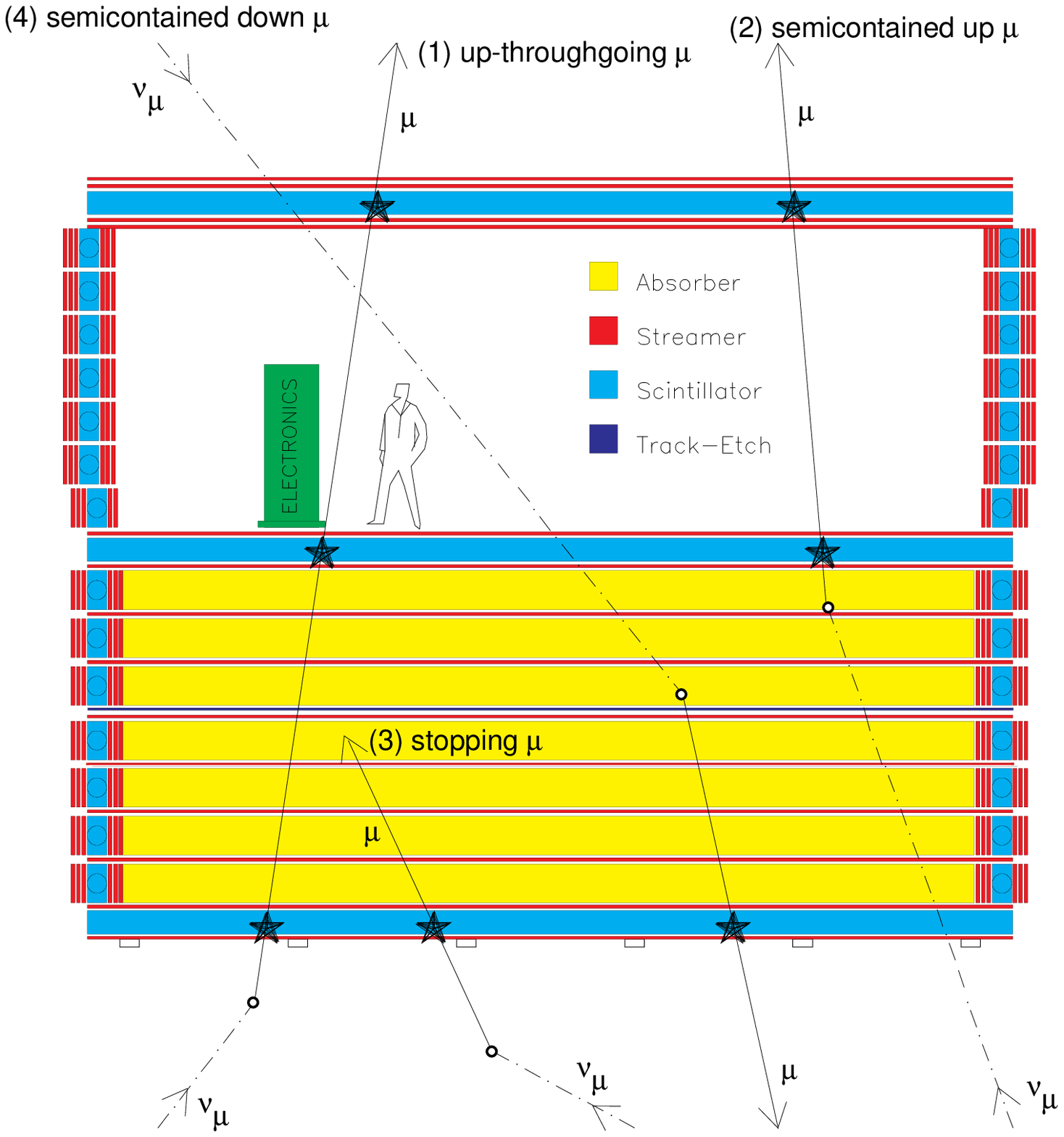}
    \hspace{0.8cm}
    \includegraphics[width=7.1cm]{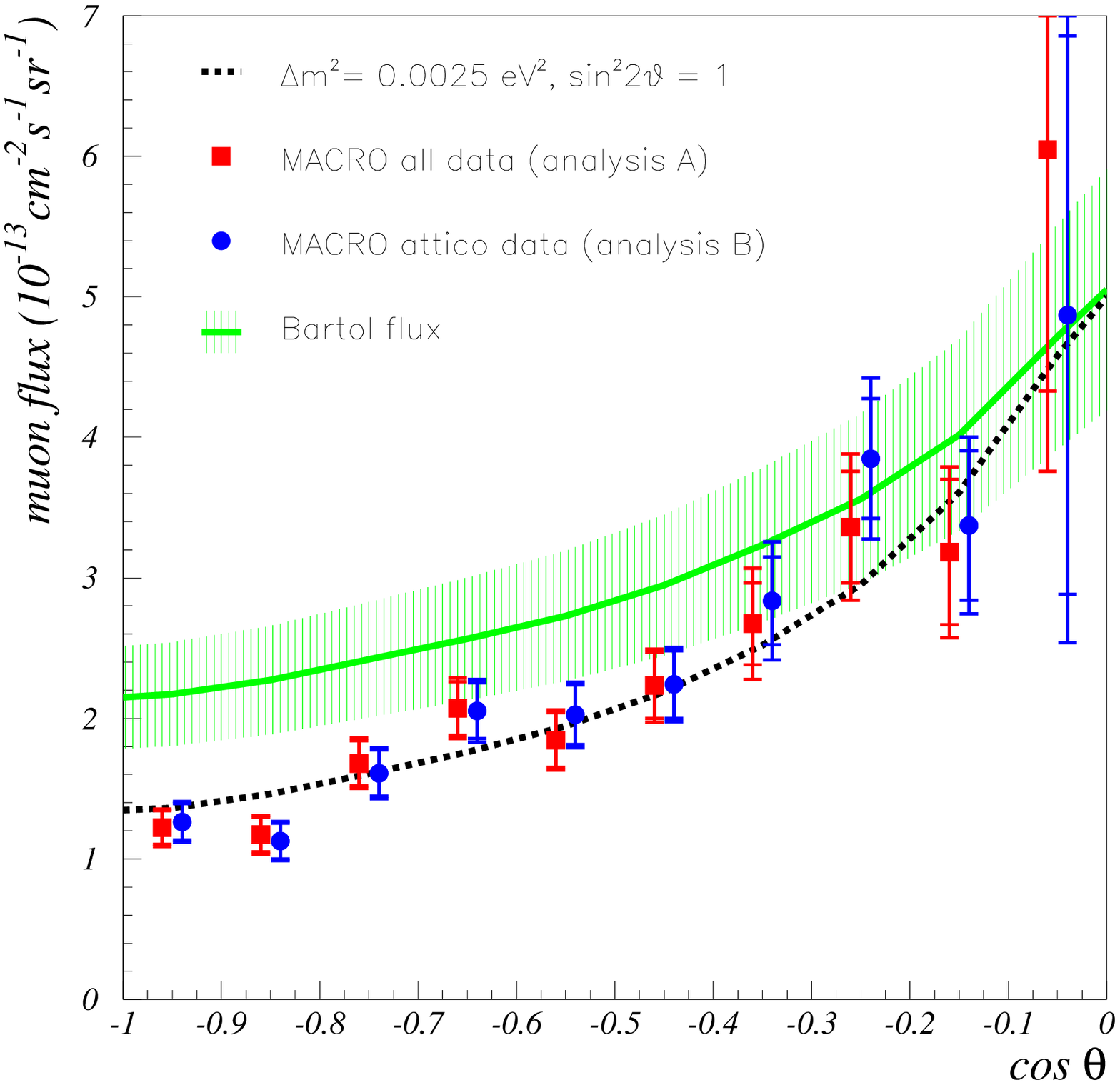}     }}
    \caption{\it 
    (A) Atmospheric $\nu$-induced event topologies.
    (B) Upward-throughgoing muon flux (the results of two different analyses are shown). 
    \label{topo} }
\end{figure}

The MACRO detector was located in the Gran Sasso underground
laboratory~\cite{artitec}.
Thanks to its large area, fine tracking granularity and up-down symmetry, 
it was a proper tool for the study of upward-travelling muons
and neutrino interactions in the apparatus. 
The different kinds of neutrino events detected by MACRO are shown in Fig.~\ref{topo}A~: 
(1)~upward-throughgoing muons, (2)~semicontained upgoing muons, (3)~upgoing stopping and 
(4)~semicontained downgoing tracks. The sample (1) 
is due to more energetic neutrinos ($<E_\nu>\ \sim 50\ GeV$)
producing muons also at long distances from the detector. 
The other samples are due to neutrinos of lower energy ($<E_\nu>\ \sim 4\ GeV$). 
The samples (3) and (4) are indistinguishable and therefore they are studied 
together (3+4). 

\begin{figure}[t]
  \vspace{-0.5cm}
  \centerline{\hbox{ 
    \hspace{-0.3cm}
    \includegraphics[width=7.0cm]{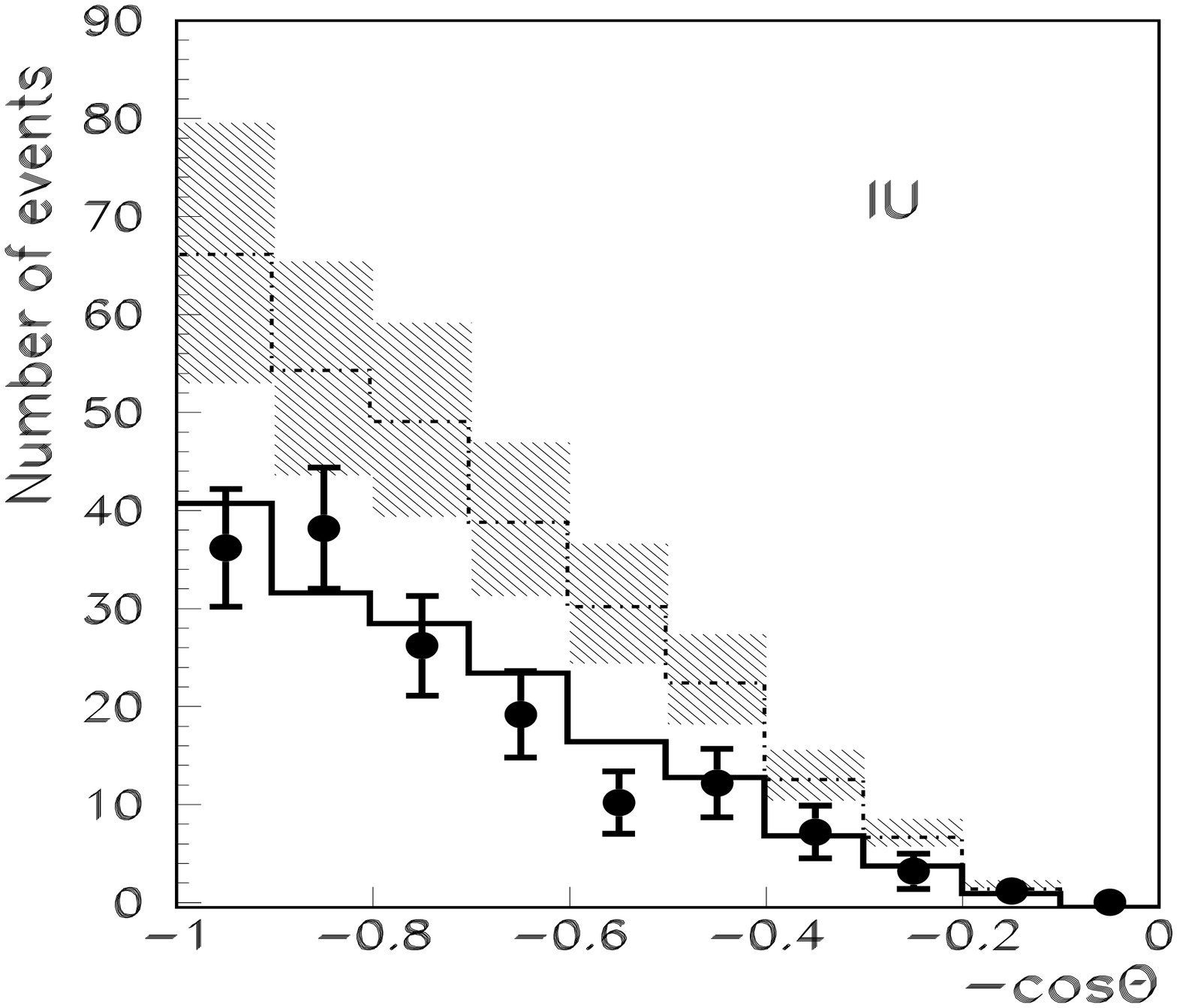}
    \hspace{0.3cm}
    \includegraphics[width=7.0cm]{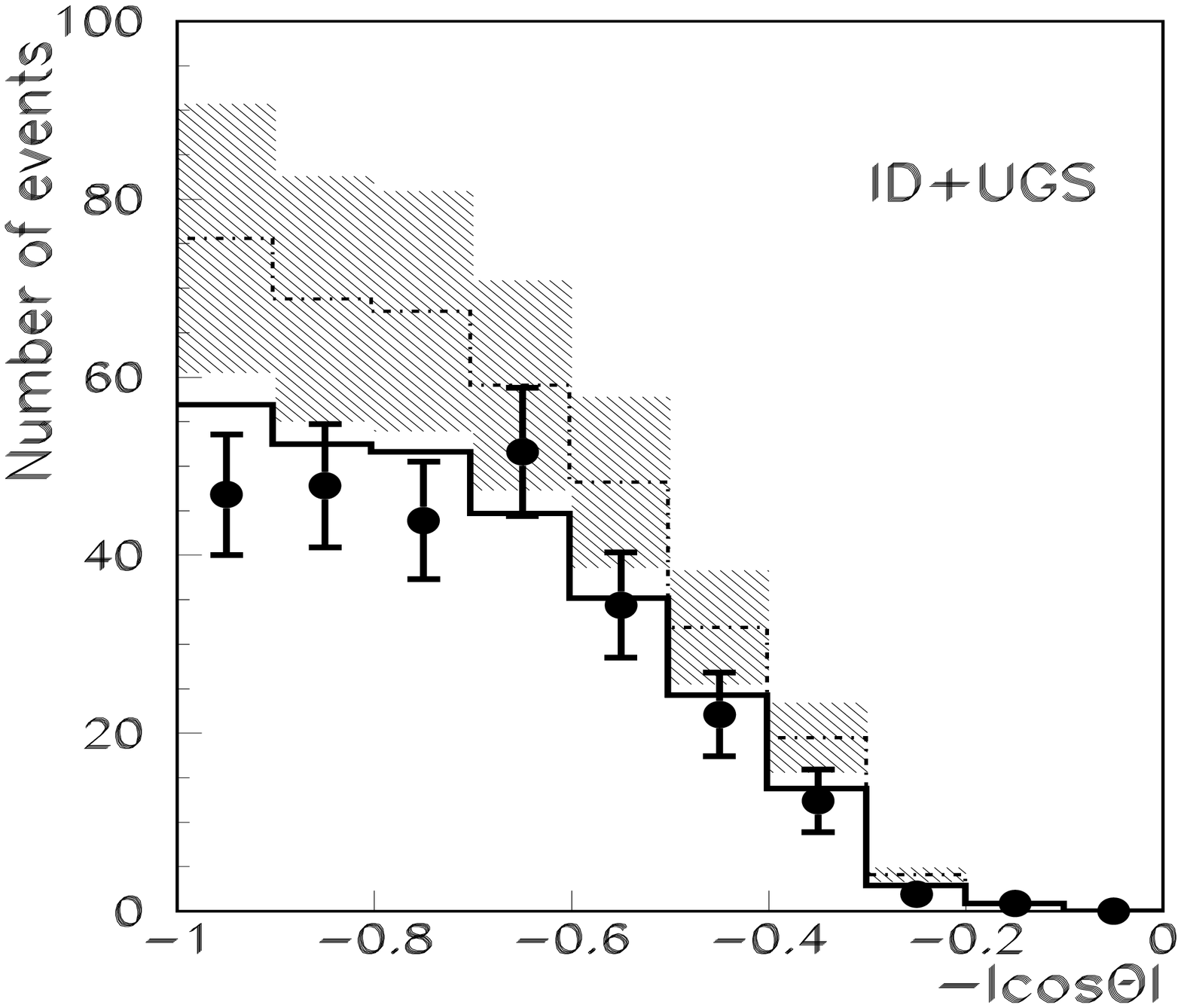}    }}
    \caption{\it Angular distributions for samples (2) and (3+4).
    \label{low} }
\end{figure}

\begin{figure}[htbp]
  \centerline{\hbox{ 
    \hspace{-0.3cm}
    \includegraphics[width=7.0cm]{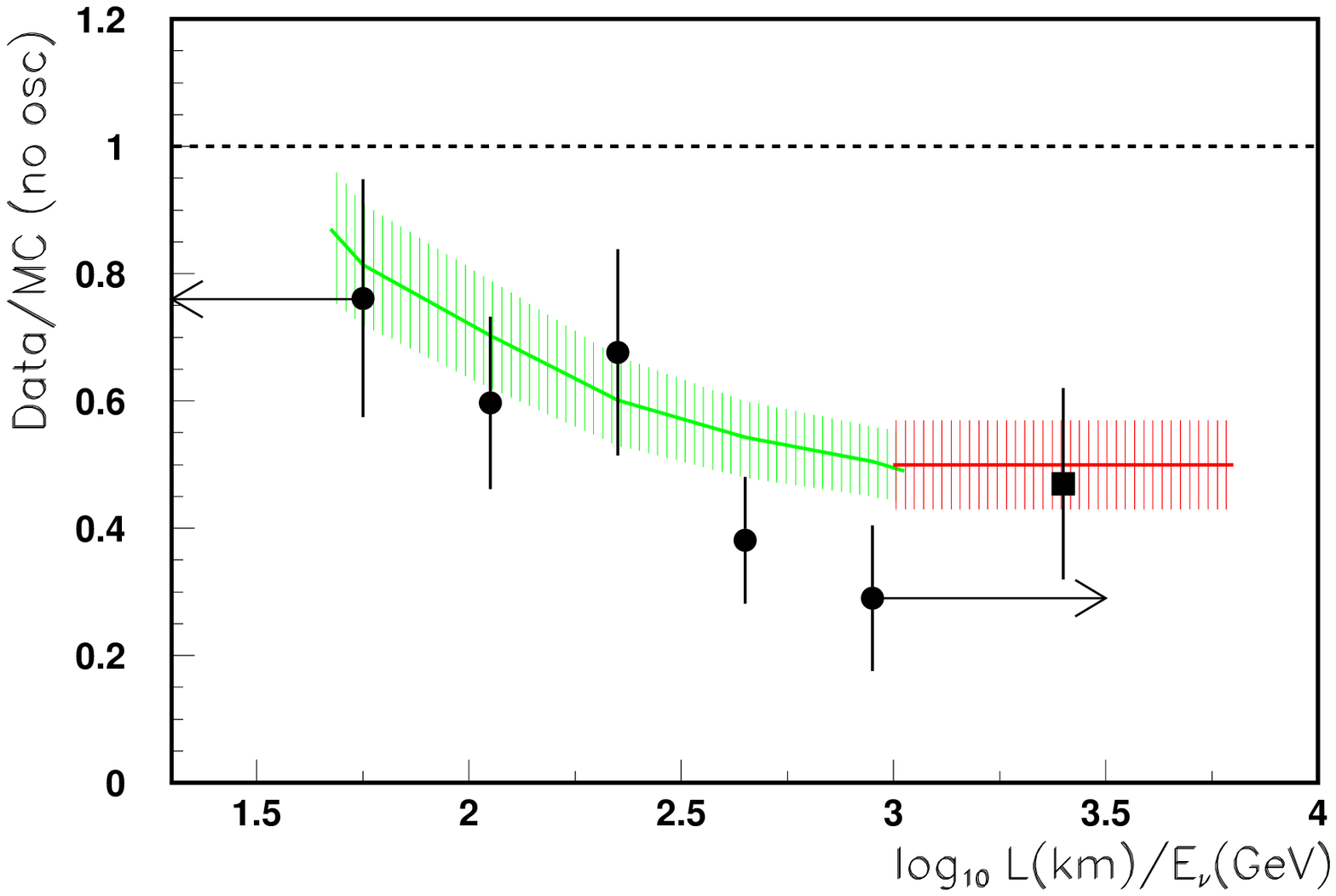}   
    \hspace{0.3cm}
    \includegraphics[width=7.0cm]{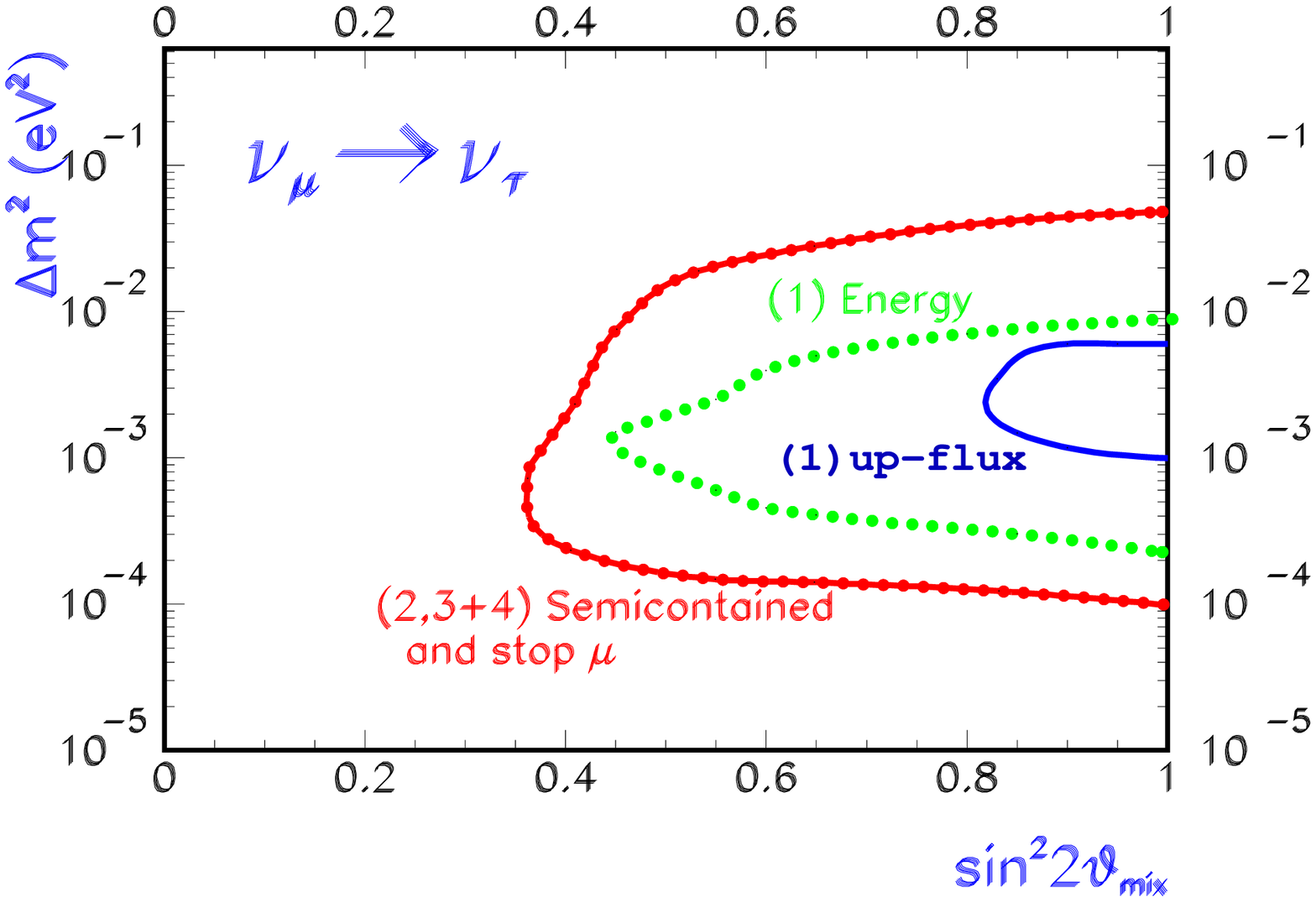}   }}
    \caption{\it 
    (A) Data on simulation ratio vs $L/E_\nu$ (energy estimate based on 
        Multiple Coulomb Scattering). The curve is obtained assuming 
	$\nu_\mu \to \nu_\tau$ oscillation. 
    (B) Allowed regions assuming $\nu_\mu \to \nu_\tau$ oscillation.
\label{mcs} }
\end{figure}

In Fig.s~\ref{topo}B and~\ref{low} the angular measurements are shown for samples (1), 
(2) and (3+4), respectively. All distributions are so far from the 
expectation assuming no oscillation. Assuming $\nu_\mu \to \nu_\tau$ oscillation the 
measurements become compatible with expectation~\cite{highnu}. The best fit parameters result 
$\sin^2 2 \theta_{mix} = 1$ and $\Delta m^2 = 0.0025 \ eV^2$. 
The measurement of the Multiple Coulomb Scattering~\cite{MCS}
permits to estimate the energy of muons in sample (1). Neutrino energies 
are inferred by means of Monte Carlo methods. In Fig.~\ref{mcs}A the
data/expectation ratio as a function of estimated $L/E_\nu$ is shown. The last 
point is due to sample (2).

In Fig.~\ref{mcs}B the 90\% C.L. allowed regions assuming $\nu_\mu \to \nu_\tau$ 
oscillation are shown.
The smaller area is estimated by normalization and angular distribution of
sample (1). The medium area is due to $\mu$-energy estimate for the same
sample. The larger area is deduced by (2) and (3+4) low energy samples.

\section{Neutrino astronomy}

A sample of 1356 upward-travelling muons ($E_\mu > 1 \ GeV$) has been studied to look for 
astrophysical signals~\cite{nuastro}. The data do not show $\nu$-excess from selected 
pointlike sources. Neutrino-induced muon-flux limits have been established at the level 
of $\sim~10^{-14}\ cm^{-2}\ s^{-1}$. Recently the microquasar GX339-4 has been proposed 
as a source of TeV neutrinos~\cite{microqu}. We note that MACRO detected 7 events 
in the direction of GX339-4, while 2.3 atmospheric neutrinos were expected~\cite{nuastro}.  
Another analysis searched for high-energy upward-travelling muons due to astrophysical
high-energy diffuse neutrinos. No statistically significant signal has been found
and an upper limit has been set on diffuse neutrino flux assuming $E_\nu^{-2}$ as 
power law spectrum~\cite{nudiffu}.

\end{document}